\documentclass[aip,rsi,reprint,nolongbibliography,superscriptaddress,floatfix]{revtex4-2}
\usepackage[T1]{fontenc}
\usepackage[utf8]{inputenc}
\usepackage{graphicx}
\usepackage{amsmath}
\usepackage{xcolor}
\usepackage{txfonts}
\usepackage{times}
\usepackage{upgreek}
\usepackage{hyperref}
\usepackage{gensymb}
\usepackage{orcidlink}
\usepackage[normalem]{ulem}

\definecolor{greyed}{RGB}{170, 170, 170}
\definecolor{myblue}{RGB}{0, 40, 140}
\definecolor{blankcolor}{RGB}{220, 220, 230}
\hypersetup{
    unicode=true,          
    pdftitle={High-power RF amplfier for ultracold atom experiments},    
    pdfauthor={P Thekkeppatt, et al.},     
    colorlinks=true,       
    linkcolor=myblue,          
    citecolor=myblue,        
    filecolor=myblue,      
    urlcolor=myblue           
}

\begin{document}
	\title{High-power RF amplifier for ultracold atom experiments}

	\author{Premjith Thekkeppatt\,\orcidlink{0000-0002-1884-8398}} 	
       \email{RFamplifier@strontiumbec.com}
 \affiliation{Van der Waals-Zeeman Institute, Institute of Physics, University of Amsterdam, Science Park 904, 1098 XH Amsterdam, The Netherlands}
    
    \author{Edwin Baaij\,\orcidlink{}}
    \affiliation{Technology Center, Faculty of Science, University of Amsterdam, Science Park 904, 1098 XH Amsterdam,
The Netherlands}
    
    \author{Tijs van Roon \,\orcidlink{}}
    \affiliation{Technology Center, Faculty of Science, University of Amsterdam, Science Park 904, 1098 XH Amsterdam,
The Netherlands}
    
	\author{Klaasjan van Druten\,\orcidlink{0000-0003-3326-9447}}	  
	\affiliation{Van der Waals-Zeeman Institute, Institute of Physics, University of Amsterdam, Science Park 904, 1098 XH Amsterdam, The Netherlands}
	
    \author{Florian Schreck\,\orcidlink{0000-0001-8225-8803}}
    \affiliation{Van der Waals-Zeeman Institute, Institute of Physics, University of Amsterdam, Science Park 904, 1098 XH Amsterdam, The Netherlands}
	    
	\date{\today}
	
\begin{abstract}
  We report on the design and characterization of a high-power amplifier with an output power of 36.5\,dBm for a frequency range of 50\,MHz to 1000\,MHz with a total gain of 40\,dB. The amplifier is optimized for driving acousto-optic and electro-optic modulators for ultracold atom experiments. This amplifier is a 19 inch rack unit, with a power efficiency of >35\%, and 0.01\,dBm long-term stability. Schematics and other design materials are publicly available under an open hardware license.

\end{abstract}

\maketitle

\section{Introduction}
Optical modulators are frequently utilized in optics laboratories to control and manipulate laser frequency, phase, and intensity. They are particularly prevalent in the complex experimental setups of atomic, molecular and optical (AMO) physics
\cite{volponi_circus_2024,shammah_open_2024}. Two widely used optical modulators are acousto-optic and electro-optic (AO and EO) modulators, which are employed for frequency modulation, intensity modulation, pulse shaping, and controlling the phase of laser beams\cite{donley_double-pass_2005}.
These modulators play a crucial role in experiments involving ultracold atoms and ions, enabling the control of various laser beams used for cooling and optical trapping, as well as to perform operations on atoms and ions using lasers. The AO/EO modulators require a high-power radio-frequency (RF) signal to operate. In most cases, these modulators are driven by a high-power RF amplifier \cite{barker_flexible_2019}, which amplifies the output of an RF source, such as a direct frequency synthesizer (DDS) or voltage-controlled oscillator (VCO), to power levels of 1\,--\,4 watt. Due to the ever increasing complexity and large number of laser beams used in ultracold atom and ion experiments for cooling, trapping and coherent control, the number of AO/EO modulators used in such a setup today can be 20 to 50 and will increase further. This requires a large number of RF amplifiers for each setup, operating at various RF frequencies especially in the 30 to 500\,MHz range. It is therefore desirable to design a compact, rack-based amplifier with good power efficiency (to reduce heat dissipation) while maintaining low-noise performance. 

The recent development of RF SoCs (system on chips) along with a new generation of high-speed digital-to-analog converters (DACs) has significantly improved RF frequency synthesis in the last decade \cite{sitaram_programmable_2021,stefanazzi_qick_2022,morzynski_open-source_2023}. There are still significant challenges to be addressed in RF amplifiers, the main one being power efficiency (typically limited to <20\% at present) \cite{barker_flexible_2019,kasprowicz_artiq_2020}. Lower power efficiency imposes significant design restrictions and requires better thermal management, enforcing the use of bulkier heat sinks and larger packaging for better cooling. RF amplifiers combining higher power efficiency with compact packaging while maintaining the performance of traditional amplifiers will play a significant role in the advancement of ultracold atom and ion experiments to a smaller form factor.

We present an open-source design of an RF amplifier with a high output power of 36.5\,dBm, a gain of 40\,dB and high power efficiency (>35\%). This RF amplifier is designed for AO/EO modulators with an average gain flatness of 1.1\,dB over a frequency range of 50 to 1000\,MHz packaged in a 19 inch rack module (FME07). 

In this paper, we discuss hardware design in Section\,\ref{sec:design} and characterization in Section\,\ref{sec:char}. We summarize the results and discuss future improvements in Section\,\ref{sec:conc}. 
  \begin{figure*}
    \centering
    \includegraphics[width=\textwidth]{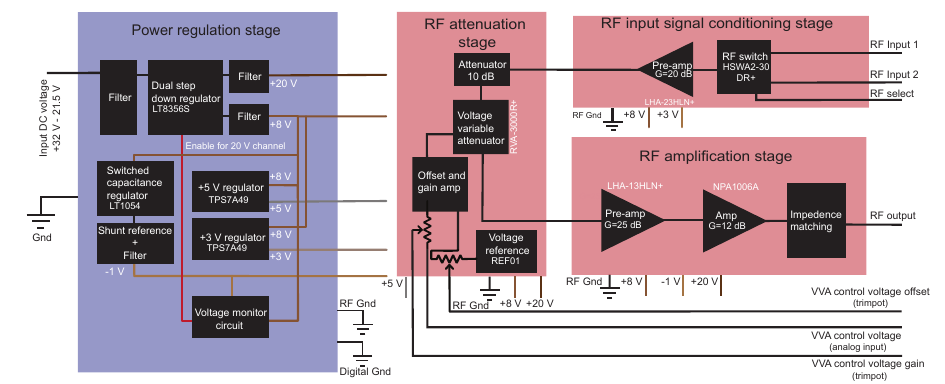}
    \caption{Schematic diagram of the RF amplifier. The amplifier consists of RF stages (right, pink rectangles) and a power regulation stage (left, violet rectangle), which includes an under-voltage lockout system. The front panel input/outputs are shown on the right.}
    \label{fig:AmpSchematic}
\end{figure*}
   
\section{Design}
\label{sec:design}

The RF amplifier is designed with the aim of driving AO/EO modulators which are used mainly to control and modulate the frequency, amplitude, and phase of laser beams. The design 
is focused on improving the power efficiency of the system while still producing high power output with long-term RF power stability. We use a Gallium Nitride (GaN) on silicon RF amplifier chip that is commonly used for wideband telecom operations. GaN HEMT (High Electron Mobility Transistor) devices offer high-frequency and high-linearity, are cost-effective, and are typically used for communication applications. In addition to the performance aspects, they offer reliability and long lifespan. The choice of this GaN-based amplifier chip plays an important role in obtaining a high power output of >35\,dBm with large bandwidth (50 - 1000\,MHz), making it ideal for driving AO/EO modulators \cite{lu_review_2023}.

As summarized in Fig.\,\ref{fig:AmpSchematic}, our design consists of four main stages, namely one power regulation stage (violet) and three RF stages (pink). The power regulation stage converts the input voltage into the voltages required by the RF stages. We employ switching regulators and obtain an efficiency of 90\%. The first RF stage pre-amplifies one of two RF inputs, which can be selected through a digital input. The second stage attenuates or intensity modulates the RF signal in dependence of an external input voltage, which is important for use in driving AO/EO modulators. The third stage amplifies the RF signal using another pre-amp and a high-power RF amplifier chip. We take special care in our component selection to obtain high-power output with a flat gain profile over the frequency range of interest. The RF amplifier is designed with a tunable negative gate voltage for the high-power RF amplifier chip, in order to have the ability to optimize the output power, power consumption, and total harmonic distortion of the RF amplifier. As a result, we achieve a highly linear performance along with low harmonic distortion, which is advantageous when performing coherence-preserving operations on atoms and ions.

\subsection{Switching power regulation unit \label{sec:level1}}
The design of the power regulation stage is based on a single positive input supply voltage in the range of +21.5 to 32\,V. We use a dual channel step-down regulator that delivers up to 2\,A of continuous current per channel (LT8653S) to convert the input voltage to +20\,V and +8\,V. This regulator regulates the entire power input of our system, as it is used to power the high-power RF amplifier chip (NPA1006, +20\,V) and the low-power, low-noise RF amplifiers (LHA-23HLN+ and LHA-13HLN+, +8\,V). The switching frequency of the regulator is tuned to 610\,kHz. The regulator outputs are filtered to suppress switching noise by -55\,dB. The switching frequency is chosen to maximize power efficiency while at the same time not leading to spurious sidebands on the RF output that would significantly influence AO/EO modulators. 

\begin{figure}[htbp]
    \centering
    \includegraphics[width=0.9\linewidth]{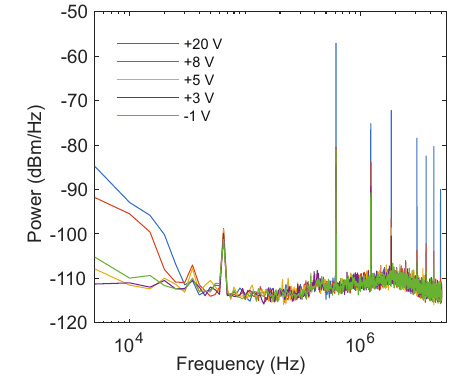}
    \caption{ Noise power spectral density of different power rails produced by the power regulation stage measured using a Rohde $\&$ Schwartz (R$\&$S) FPL1007, +20\,V (blue), +8\,V (red), +5\,V (yellow), +3\,V (violet), -1\,V (green) for an input voltage of +26\,V.}
    \label{PSUcharac}
\end{figure}

We use a low-dropout, low-noise, linear voltage regulator (TPS7A49) to convert the +8\,V channel of the switching regulator into +3\,V and +5\,V. These additional voltage supplies (+3\,V, +5\,V) are for the RF switch and the voltage-controlled RF power attenuator (VVA). We use a switched-capacitor voltage regulator (LT1054) in combination with a precision shunt voltage reference (LT1634) and a filtering stage to produce a negative voltage of $\sim$\,$-2.5$\,V. This voltage is divided and buffered with an op-amp to produce a stable and tunable negative gate voltage for the high-power RF amplifier chip. Creating a negative power rail from a positive rail using a switched-capacitor voltage regulator helps us avoid the use of a negative input rail, thereby simplifying the use of the RF amplifier. We operate the switched-capacitor voltage regulator at 25\,kHz, and the LT1634 chip and filtering stage offer a 55\,dB suppression at 25\,kHz. We observe an integrated noise spectral density of $<$\,-\,93\,dBm/Hz in all switched-mode regulated power rails. A measurement of the noise spectral density of different power rails is shown in Fig. \ref{PSUcharac}. Automatic shutdown of the power regulation stage happens when the input voltage is less than 21.5\,V.

\begin{figure}[htbp]
    \centering
    \includegraphics[width=0.9\linewidth]{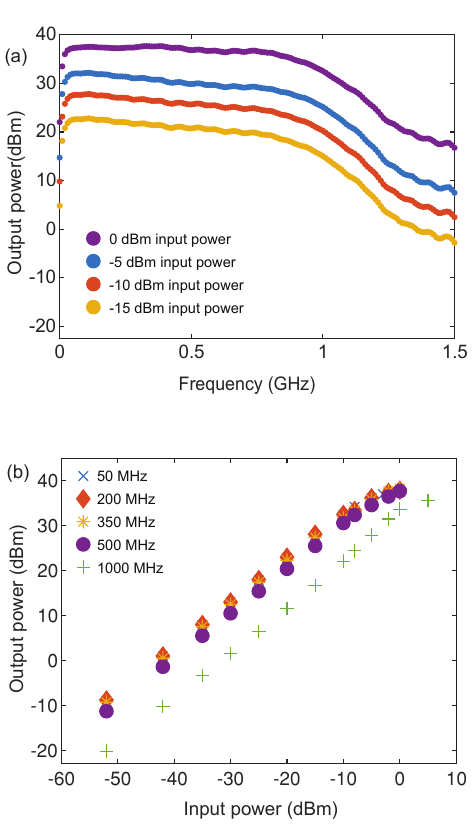}
    \caption{(a) Measurement of gain flatness of the RF amplifier over the range of 10\,MHz to 1.5\,GHz using a R$\&$S FPL1007 spectrum analyser for different input powers, -15\,dBm (yellow), -10\,dBm (red), -5\,dBm (blue), 0\,dBm (violet). (b) Measurement of input vs. output power for various frequencies (see legend). }
    \label{gainfatness}
\end{figure}

\subsection{RF input signal conditioning stage  \label{sec:level2}}
    The RF input signal conditioning stage consists of an absorptive RF switch and a low-noise pre-amplifier. The RF input is 50\,$\Omega$ impedance-matched for frequencies from 20\,MHz to 1 \,GHz. Our design supports the use of two RF input signals, selected using a dual-channel RF switch (HSWA2-30DR+) \cite{minicircuits_HSWA2-30DR+}. The RF switch also helps us switch on or off the RF power output in $\sim$2$\,\mu$s and enables the use of different RF sources as input.
    By default, our design uses RF input 1 if 0\,V is applied to the digital input channel select port. The use of an RF switch in addition to the analog attenuator described below enables a total attenuation of >60\,dB, which significantly helps to avoid RF power leaks to the high-power stage. The RF signal selected by the switch is sent to a pre-amplification stage consisting of a low-noise amplifier (Mini-Circuits LHA-23HLN+) \cite{MinicircuitsLHN-23LN+}. This amplification stage offers a gain of 20\,dB with a very low noise figure of <1.23\,dB. 
    
\subsection{RF attenuator stage  \label{sec:level3}}
    The pre-amplified RF input is sent to an RF attenuator stage. This stage consists of a voltage variable attenuator (VVA, Mini-Circuits RVA-3000R+), which offers an input frequency bandwidth of 20 -- 3000\,MHz, low signal distortion and minimal phase deviation. This voltage-controlled RF attenuator allows us to reduce the input RF power, thereby controlling the output RF power, with a modulation bandwidth of 
    $\sim$\,$200\,$kHz \cite{minicircuits_rva-3000r_nodate}. This stage enables control of the RF power through an analog input voltage, and thereby also opens up the possibility of power stabilization (e.g. optical power via an AO modulator) using a PID loop. The control voltage sent to the VAA is the sum of an external VAA control voltage and a trimpot-tunable VAA control voltage offset derived from a voltage reference (REF01). Before summation, the VAA control voltage passes a voltage divider trimpot, which enables the user to adjust the ratio of external control voltage to RF attenuation (i.e. the VAA control voltage gain). The VAA control voltage offset allows the user to adjust the minimum gain of the RF amplifier. This feature is in particular useful if no external VVA control voltage is applied, e.g. to adjust the RF power to the optimal power required for an AOM.
        
\begin{figure}[htbp]
    \centering
    \includegraphics[width=1\linewidth]{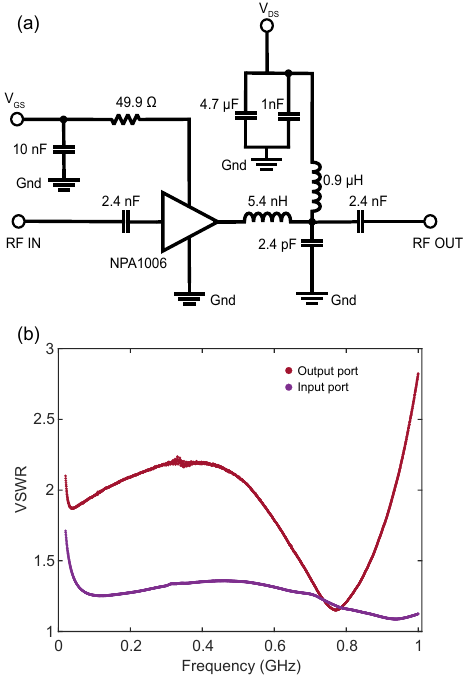}
    \caption{(a) Schematic of the output impedance matching network for the NPA1006 GaN high-power RF amplifier. The circuit transforms the amplifier's output impedance to 50 $\Omega$ across the operational frequency range ($50-1000$\,MHz) using a combination of inductors, capacitors, and resistors. The input gate voltage (V$_{\rm GS}$) and drain voltage (V$_{\rm DS}$) connections are shown with appropriate decoupling capacitors for RF isolation. (b) Voltage Standing Wave Ratio (VSWR) measurements of the input (blue) and output (red) ports of the RF amplifier as a function of frequency, demonstrating the effectiveness of the impedance matching.
    }
    \label{vswr}
\end{figure}

\subsection{RF amplification stage \label{sec:level4}}
   This stage consists of a pre-amplifier and a high-power RF amplifier chip, offering a total gain of 37\,dB and output power of 36.5\,dBm. We use a low-noise amplifier (Mini-Circuits LHA-13HLN+) as the pre-amplifier, which offers a gain of 25\,dB, yielding a maximum output power of 25\,dBm with a noise figure of <1.4\,dB \cite{minicircuits_LHA13HLN-1}. The $50\,\Omega$ impedance-matched design of the LHA-13HLN+ makes it a good choice, because it reduces the complexity of custom-designed impedance-matching circuits.
   
    We use a Macom NPA1006 GaN on silicon device as high-power RF amplifier chip \cite{macom_gan_nodate}. It can deliver output powers up to 41\,dBm in a frequency range of 50 - 1000\,MHz. Using RF amplifier chips made of Gallium Nitride (GaN) on silicon substrate offers several significant advantages, such as higher efficiency, higher power handling capacity, and better thermal performance. The NPA1006 offers a typical gain of 12\,dB, a power efficiency of $\sim$\,$60\%$ and a $50\,\Omega$ input.  
    To ensure efficient power transfer and minimize signal reflections, we designed a discrete output impedance matching network for the NPA1006 high-power RF amplifier chip to transform the amplifier’s output impedance to $50\,\Omega$ across the operational frequency range. The schematic of this matching circuit is shown in Fig.4(a). The performance of this impedance matching circuit is demonstrated in Fig. 4(b), where we present measurements of the Voltage Standing Wave Ratio (VSWR) at both the input and output ports of the amplifier. The low VSWR values observed over the frequency range confirm successful impedance matching and efficient RF power delivery.

    We have extensively characterized several high-power RF amplifier chips (NPA1006) by operating them at various drain and gate voltages. We then optimized those voltages to obtain high efficiency, amplification and linear performance for an output power of 36.5\,dBm. In our design, the drain voltage of the amplifier is reduced from the maximum allowed value of 28\,V to 20\,V, reducing our total output power from 41\,dBm to 36.5\,dBm. We also observe an optimum negative gate voltage of $\sim$\,-1\,V for maximum efficiency, output power, and low harmonic distortion. To achieve the best performance, this voltage must be individually tuned for each RF amplifier in a range of $1\pm0.15$\,V and a precision of $\sim$\,0.02\,V.
    
 \begin{figure}[htbp]
    \centering
    \includegraphics[width=0.9\linewidth]{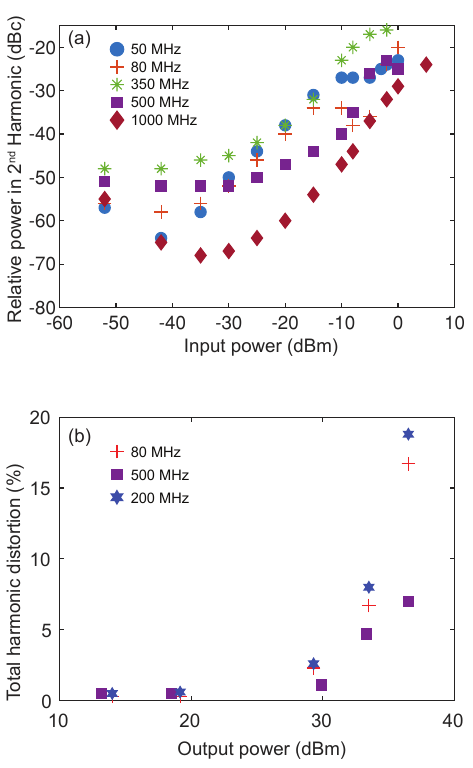}
    \caption{Total harmonic distortion measurement. (a) Relative output power in the second harmonic component (at twice the frequency of the input), as a function of input power. (b) Total harmonic distortion considering the first five harmonic components over output power at the fundamental frequency.}
    \label{THD}
\end{figure}

\subsection{Power sequencing system}
 From our extensive testing of the NPA1006 chip, we realized that it is important to apply the negative gate voltage before the drain voltage to protect the NPA1006 chip from damage. This was motivated by the observation of an increased failure rate of the NPA1006 chip in the absence of power sequencing. We therefore implement a power sequencing circuit, as recommended by the NPA1006 datasheet \cite{NPA1006}. The power sequencing circuit monitors the negative gate voltage and only when this voltage settles to -1\,V, an enable signal is produced that soft starts the +20\,V channel, which supplies the drain voltage of the NPA1006 chip. To implement this, we used the +8\,V output of the LT8653S to produce the gate voltage using the switched-capacitor voltage regulator. A voltage monitor circuit is used to detect and compare the gate voltage, and only when the gate voltage settles to -1\,V, a delayed enable output signal is produced, which then executes a soft start of the +20\,V channel, thereby controlling the +20\,V drain voltage for the NPA1006. A delay time of 50\,ms is applied before enabling the drain voltage output. This power sequencing stage always ensures that the drain voltage follows the gate voltage for this NPA1006, reducing the risk of damage. Along with sequencing, we also implement an under-voltage lockout system of the + 8\,V power rail with a threshold at 7.5\,V. 
 
\subsection{RF design considerations and thermal management}
 We use a 4-layer printed circuit board (PCB) made from FR4 material. The 4-layer PCB design with carefully chosen layer spacing offers a 50\,\(\Omega\) impedance matching for the entire target frequency range. The choice of a 4-layer PCB also offers better thermal conductivity and heat dissipation compared to a 2-layer board.
 Even with its high power efficiency, the RF amplifier dissipates 6\,W at maximum output power. To efficiently channel this heat away from the high-power RF amplifier chip and keep the temperature always below the maximum operating temperature, a carefully designed thermal management system is required. We do not use specific heatsinks for thermal management. Instead, we use an RF shielding case (Laird BMI-S-207), which is soldered to ground pads close to the amplifier stage. In addition to offering EMI shielding, it also improves heat dissipation. Together, the 4-layer PCB design and the RF shielding allow us to perform passive air cooling efficiently. We observe a steady-state temperature of 80\degree{}C of the NPA1006 chip without forced air cooling for maximum output power. For normal operation, these amplifiers are placed in a 19 inch subrack (BGT 384/180IZ) equipped with three axial AC fans on their bottom. We obtain an air flow of 7.8\,\(\text{m}^3\)/min at maximum fan power. Forced air cooling reduces the NPA1006's temperature to 65\,\degree{}C. This temperature is well below the absolute maximum rating for the high-power RF amplifier chip temperature of 85\degree{}C. 
 \begin{figure*}[htbp]
    \centering
    \includegraphics[width=\textwidth]{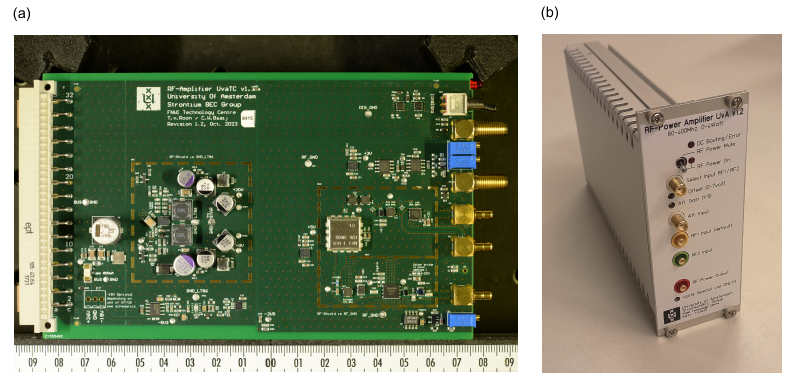}
    \caption{(a) Photograph of the high-power RF amplifier PCB, showing the layout of key components including the power regulation, RF input conditioning, attenuation, and amplification stages. (b) Image of the fully assembled RF amplifier housed in a 19-inch rack-mountable enclosure, featuring accessible front-panel controls for RF input selection, attenuation adjustment, and output monitoring. The compact design integrates all functional modules necessary for driving acousto-optic and electro-optic modulators in ultracold atom experiments, demonstrating robust construction suitable for laboratory deployment.}
    \label{PCB and Casing}
\end{figure*}
   \begin{table}[htbp]
    \begin{tabular}{lcc}
    \hline
        Parameter & Typ  & Unit  \\
        \hline
        Frequency range & 50 - 1000 & MHz \\
         Gain & 40 & dB \\
         Gain flatness & 1.1 & dB \\
         Output power & 36.5 & dBm \\
         Output power stability& 0.01 & dBm \\
         Output third-order intercept point & 52 & dBm \\
         Power efficiency & >35 & \% \\
         RF isolation & >100 & dB \\
         Input supply voltage & 32 - 21.5 & V \\
         Input supply current & 400  & mA \\
         
    \end{tabular}
    \caption{Key parameters and their measured values for the RF amplifier }
    \label{tab:paramters}
\end{table}

\begin{figure}[htbp]
    \centering
    \includegraphics[width=0.9\linewidth]{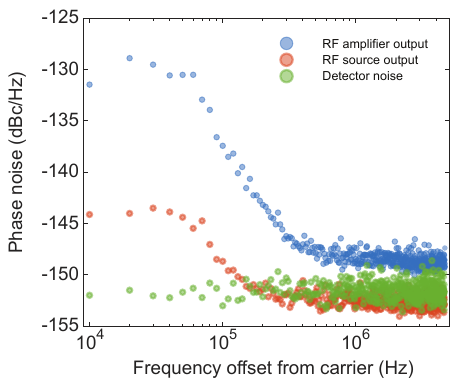}
    \caption{Added phase noise measurement of RF amplifier at an output power of 36.5 dBm measured with R$\&$S SMCV100B RF source and R$\&$S FPL1007 spectrum analyser.}
    \label{phasenoise}
\end{figure}

 \section{Characterisation}
 \label{sec:char}
We characterize the gain flatness of the RF amplifier using a Rohde $\&$ Schwarz (R$\&$S) FPL1007 spectrum analyzer. We use the tracking generator of the R$\&$S FPL1007 as RF source for our amplifier and measure the gain for various input powers while sweeping the RF frequency. A maximum gain of 40\,dB is measured with an average gain flatness of 1.1\,dB (standard deviation) over a range of 50 -- 1000\,MHz (Fig.\,\ref{gainfatness}a). The output RF power for various input RF powers is measured at different frequencies ranging from 50 to 1000\,MHz to characterize the first-order intercept point (IP1). The behaviour observed (Fig.\,\ref{gainfatness}b) is very close to linear for all investigated frequencies up to $\sim$36\,dBm output power. Above $\sim$36\,dBm output power, saturation sets in after $\sim$\,36\,dBm, showing that the amplifiers are extremely linear up to this output power over a wide frequency range. To further investigate the linearity of the RF amplifier, we determine the third-order intercept point (IP3) by sending two closely spaced RF frequencies $f_1$ and $f_2$ to the RF input. IP3 is an imaginary point where the power of the fundamental output signals at $f_1$ and $f_2$ and the third-order intermodulation product ($2f_1-f_2$ and $2f_2 - f_1$) would intersect if the output power of both continued to increase linearly \cite{termsdefined}. To determine IP3, we use two phase-locked RF sources (R$\&$S SMCV100B and R$\&$S SMB100B) at a central frequency of 200\,MHz and with an offset of 1\,MHz between the two sources. The output power of the two RF sources is carefully calibrated to be equal. We measure the amplified output power in the fundamental frequency components and the third-order intermodulation products at various input powers, and then extrapolate the data to extract IP3. We find IP3 at a total output power of 52\,dBm. 

To characterize the total harmonic distortion (THD) of the RF amplifier, we measure the ratio of the output power of the second-harmonic component to the fundamental frequency for various input powers (Fig.\ref{THD}). For a maximum power output of 36.5\,dBm, we observe the second-harmonic component at $\sim$\,20\,dBc. We also characterize the harmonic distortion due to third and fourth harmonic components, finding it to be $>$\,-\,25\,dBc for the frequency range of interest. THD can be further reduced for a particular frequency and power to $\sim$\,-40\,dBc or less by optimizing the negative gate voltage. We also measure the phase noise introduced by the RF amplifier at a maximum output power of 36.5\,dBm. To characterize the added phase noise from the RF amplifier, we use a spectrum analyzer (R$\&$S FPL1007) and an RF generator (R$\&$S SMCV100B) in a phase-locked configuration. In Fig.\ref{phasenoise}, for a carrier frequency of 200\,MHz, we show the measured phase noise of the carrier as a function of offset frequency. A maximum of 15\,dBc/Hz increase in phase noise of the RF signal is observed for an output power of 36.5\,dBm. 

The long-term power stability is characterized using a calibrated RF power meter (R$\&$S NRP-ZAD1). In steady-state operation, for an output power of 36.5\,dBm, we measure a standard deviation of the output power of 0.01\,dB over a time period of 30 minutes. To characterize the rise time and fall time of the VVA and RF switch, an arbitrary waveform generator (Rigol DG4102) is used to synthesize a square waveform at various frequencies as a control signal and the RF amplitude response is measured using the RF power meter mode of the R$\&$S FPL1007. A rise time of 5\,$\mu$s and a fall time of 2.5\,$\mu$s are measured. The RF switch switching time is measured to be $\sim$\,3\,\(\mu\)s. This fast switching enables the production of short pulses with precise timing for laser cooling, trapping, and quantum state manipulation (e.g., Rabi oscillations, Ramsey interferometry).

A direct comparison with the Mini-Circuits ZHL-5W-202-S+ amplifier highlights the advantages and unique features of our design. The ZHL-5W-202-S+ is a connectorized, high-power, linear amplifier that delivers up to 5\,W saturated output power (approximately 37\,dBm) across a wide frequency range of 10 to 2000\,MHz, with a typical gain of 50\,dB and gain flatness of ±2\,dB. Our amplifier achieves a comparable maximum output power of 36.5\,dBm over the 50–1000\,MHz range, with a total gain of 40\,dB and superior gain flatness of 1.1\,dB (standard deviation). While the ZHL-5W-202-S+ offers a broader frequency range and higher gain, our design is optimized for the specific requirements of acousto-optic and electro-optic modulators in ultracold atom experiments, emphasizing power efficiency (>35$\%$), and compact rack integration. Importantly, our amplifier includes both a VVA and an RF switch and demonstrates fast switching times (3 $\mu $s). This is important for time-critical quantum control and feedback applications. Additionally, our open hardware approach enables full transparency, customization, and cost-effective scalability,  which is particularly relevant as the number of modulators in modern AMO experiments continues to grow.

\section{Conclusion}
\label{sec:conc}
In summary, we have designed a high-power RF amplifier that can produce a maximum output power of 36.5\,dBm for frequencies of 50 to 1000\,MHz with 50\,\(\Omega\) RF in- and outputs, which is well suited to drive AO/EO modulators. The amplifier has a 19 inch rack-compatible design with a power efficiency of >35\% at max output power, and works with a single positive rail input between 21.5 and 32\,V. The amplifier is equipped with an RF switch and a voltage variable attenuator for high-bandwidth RF amplitude modulation. We show an extensive characterization of various aspects of the amplifier performance, including phase noise, RF power noise, gain flatness, and linearity. Our design is permissively licensed,\cite{CERN_OHL} making it easily possible for other parties to adopt and improve the design. 
  
\begin{acknowledgments}
    We gratefully  acknowledge support from the Technology Centre of the University of Amsterdam. We thank A. Sitaram for a critical reading of the manuscript. This work was financially supported by NWO (Nederlandse Organisatie voor Wetenschappelijk Onderzoek, programme grant No. 680.92.18.05, Quantum Simulation 2.0) and the Dutch National Growth Fund (NGF), as part of the Quantum Delta NL programme.  
 \end{acknowledgments}

\section*{Author contributions}

    {\bf Premjith Thekkeppatt}: conceptualization (lead), investigation (equal), resources (lead), data curation (lead), supervision (lead), writing - original draft (lead), writing – review and editing (equal)
    {\bf Edwin Baaij}: conceptualization (supporting), investigation (equal), resources (equal), writing – review and editing (equal)
    {\bf Tijs van Roon}: conceptualization (supporting), investigation (equal), resources (equal), writing – review and editing (equal)
    {\bf Klaasjan van Druten}: supervision (supporting), writing - original draft (supporting), writing – review and editing (equal), funding acquisition	(supporting)
    {\bf Florian Schreck}: supervision (supporting), writing - original draft (supporting), writing – review and editing (equal), funding acquisition	(lead)

\section*{Conflict of interest}
The authors have no conflicts to disclose.

\section*{Data availability}
The design files are openly available in the \href{https://github.com/StrontiumGroup/GaN-RF-amplifier-}{Github} repository of the StrontiumBEC Group.
\section{References}
\bibliography{RFAmp}
\end{document}